\documentclass[conference]{IEEEtran}
\IEEEoverridecommandlockouts

\usepackage{acronym}
\usepackage{amsfonts}
\usepackage[dvips]{graphicx}
\usepackage{times}
\usepackage{cite}
\usepackage{amsmath}
\usepackage{array}
\usepackage{amssymb}
\usepackage{stfloats}
\usepackage{diagbox}
\usepackage{graphicx}
\usepackage{footnote}
\usepackage{amsthm}
\usepackage{booktabs}
\usepackage{array}
\usepackage[ruled,vlined]{algorithm2e}
\usepackage{subeqnarray}
\usepackage{cases}
\usepackage{threeparttable}
\usepackage{color}
\usepackage{epstopdf}
\usepackage{multirow}
\usepackage{tabularx}
\usepackage{enumerate}
\usepackage{multicol}
\usepackage{hyperref}
\usepackage{subfig}
\usepackage{caption}

\newtheorem{proposition}{Proposition}

\def\bb0{{\mathbb{0}}}

\def\bb{{\mathbf{b}}}

\def\bh{{\mathbf{h}}}

\def\bn{{\mathbf{n}}}

\def\bv{{\mathbf{v}}}

\def\bx{{\mathbf{x}}}
\def\by{{\mathbf{y}}}

\def\b0{{\mathbf{0}}}

\def\bD{{\mathbf{D}}}

\def\bH{{\mathbf{H}}}
\def\bI{{\mathbf{I}}}

\def\bR{{\mathbf{R}}}

\def\bbC{{\mathbb{C}}}

\def\sf0{{\mathsf{0}}}

\def\rmA{\mathrm{A}}
\def\rmB{\mathrm{B}}

\def\rmD{\mathrm{D}}

\def\rm0{{\mathrm{0}}}

\acrodef{CSI}[CSI]{channel state information}
\acrodef{CSIT}[CSIT]{channel state information at the transmitter}
\acrodef{CSIR}[CSIR]{channel state information at the receiver}
\acrodef{MIMO}[MIMO]{multiple-input multiple-output}
\acrodef{SISO}[SISO]{single-input single-output}
\acrodef{MISO}[MISO]{multiple-input single-output}
\acrodef{SIMO}[SIMO]{single-input multiple-output}
\acrodef{ADCs}[ADCs]{analog-to-digital convertors}
\acrodef{SNR}[SNR]{signal-to-noise ratio}
\acrodef{AWGN}[AWGN]{additive white Gaussian noise}
\acrodef{MRT}[MRT]{maximal ratio transmission}
\acrodef{DFT}[DFT]{Discrete Fourier Transform}
\acrodef{ULA}[ULA]{uniform linear array}
\acrodef{UPA}[UPA]{uniform planar array}
\acrodef{LS}[LS]{least squares}
\acrodef{ALMMSE}[ALMMSE]{approximate linear minimum mean squared error}
\acrodef{QIHT}[QIHT]{quantized iterative hard thresholding}
\acrodef{QIST}[QIST]{quantized iterative soft thresholding}
\acrodef{SVD}[SVD]{singular value decomposition}

\ifCLASSINFOpdf

\else
\fi
\begin{document}

\title{Distributed Expectation Propagation Detection for Cell-Free Massive MIMO}
\author{
\IEEEauthorblockN{Hengtao He\IEEEauthorrefmark{1}, Hanqing Wang\IEEEauthorrefmark{2}, Xianghao Yu\IEEEauthorrefmark{1}, Jun Zhang\IEEEauthorrefmark{3}, S.H. Song\IEEEauthorrefmark{1}, and Khaled B. Letaief\IEEEauthorrefmark{1}}

\IEEEauthorblockA{\IEEEauthorrefmark{1}Department of ECE, The
Hong Kong University of  Science and  Technology,
\\Kowloon, Hong Kong, E-mail: \{eehthe, eeyu, eeshsong, eekhaled\}@ust.hk}

\IEEEauthorblockA{\IEEEauthorrefmark{2}Huawei Technologies Company Ltd., Shanghai 201206, China, E-mail: wanghanqing3@Huawei.com}

\IEEEauthorblockA{\IEEEauthorrefmark{3}Department of EIE, The Hong Kong Polytechnic University, Hong Kong,
E-mail: jun-eie.zhang@polyu.edu.hk}

\thanks{This work is supported by the Hong Kong Research Grant Council under Grant No. 16212120.}
}

\maketitle

\begin{abstract}
In cell-free massive MIMO networks, an efficient distributed detection algorithm is of significant importance.
In this paper, we propose a distributed expectation propagation (EP) detector for cell-free massive MIMO. The detector is composed of two modules, a nonlinear module at the central processing unit (CPU) and a linear module at the access point (AP). The turbo principle in iterative decoding is utilized to compute and pass the extrinsic information between modules. An analytical framework is then provided to characterize the asymptotic performance of the proposed EP detector with a large number of antennas. Simulation results will show that the proposed method outperforms the distributed detectors in terms of bit-error rate.
\end{abstract}


%
\IEEEpeerreviewmaketitle

\section{Introduction}
Massive multiple-input multiple-output (MIMO) systems have been regarded as a key enabling technology for $5$G because of its high spectral efficiency (SE), energy efficiency and link reliability \cite{massiveMIMO}. However, the SE gain becomes marginal for the cell-edge user equipments (UEs). To address this problem, a novel network architecture, namely cell-free massive MIMO, was proposed \cite{cell_free, cell_free_Zhang}. This is a disruptive emerging technology which has been recognized as a crucial and core technology for the upcoming beyond 5G and 6G networks\cite{6G}.  Such technology is expected to bring important benefits, including huge data throughput, ultra-low latency, ultra-high reliability, a huge increase in mobile energy efficiency, and ubiquitous uniform coverage. The fundamental idea is to deploy a large number of distributed  access points (APs) connected to a central processing unit (CPU) to serve  all the distributed UEs in a wide area. In particular, each AP serves all UEs in the same time-frequency resource block via a time-division duplex (TDD) mode. Compared to conventional colocated massive MIMO, cell-free networks offer  more uniform connectivity for all UEs thanks to the macro-diversity gain obtained from the distributed antennas. However, the assumption that each AP serves all UEs renders the system not scalable and incurs huge power and computational resource consumption for decoding, especially at UEs with low signal-to-interference-noise-ratios. To tackle the scalability issue, a \emph{user-centric} dynamic cooperation clustering (DCC) scheme \cite{DCC} was introduced \cite{scalable}, where each user is only served by a subset of APs.

An efficient data detection algorithm is highly desired in large-scale and complex networks, such as cell-free massive MIMO. In this aspect, some early attempts were made on centralized algorithms where the detection is totally implemented at the CPU with the received pilots and data signals reported from all APs \cite{cell_free, LSFD}. However, the computational overhead of such a centralized detection scheme is prohibitively high when the network size becomes large. To address this challenge, distributed detectors have been recently investigated. In \cite{ cell_free_distributed_receiver}, one centralized and three distributed receivers with different levels of cooperation among APs were compared in terms of SE. Unfortunately, the distributed receivers investigated in \cite{cell_free_distributed_receiver} are linear receivers and therefore highly suboptimal in terms of the bit-error rate (BER) performance. Therefore, it is of great importance to develop a distributed and \emph{non-linear} receiver to achieve a better BER performance.

In this paper,  we propose a non-linear detector for cell-free massive MIMO networks, which
is derived based upon the EP principle \cite{EP_principle} with a distributed approach \cite{Decentralized-AMP, EP_distributed}. Specifically, by adopting the linear minimum mean-square error (MMSE) estimator, the APs first detect the symbols with the local channel state information and transfer the posterior mean and variance estimates to the CPU. Then, the extrinsic information for each AP is computed and integrated at the CPU by utilizing maximum-ratio combining (MRC). Subsequently, the CPU uses the posterior mean estimator to refine the detection and the extrinsic information is transferred to each AP from the CPU via the fronthaul. Compared to other distributed linear detectors, the proposed distributed EP detector improves the detection performance by introducing more computation overhead mainly at the computationally powerful CPU. Simulation results will demonstrate that the proposed method outperforms existing distributed detectors and even the centralized MMSE detector in terms of BER.

\emph{Notations}---For any matrix $\mathbf{A}$, $\mathbf{A}^{H}$ and ${\mathrm{tr}}(\mathbf{A})$ denote the conjugate transpose and  trace of $\mathbf{A}$, respectively. In addition, $\mathbf{I}$ is the identity matrix and $\mathbf{0}$ is the zero matrix.
We use $\rmD z$ to denote the real Gaussian integration measure. That is,
\begin{equation*}
  \rmD z=\phi(z)dz, \quad \mathrm{where} \quad  \phi(z)\triangleq\frac{1}{\sqrt{2\pi}}e^{-\frac{z^{2}}{2}}.
\end{equation*}
A complex Gaussian distribution with mean $\boldsymbol{\mu}$ and covariance $\boldsymbol{\Omega}$ can be described by the probability density function,
\begin{equation*}
  \mathcal{N}_{\mathbb{C}}(\mathbf{z};\boldsymbol{\mu},\boldsymbol{\Omega})=\frac{1}{\mathrm{det}(\pi \boldsymbol{\Omega})}
  e^{-(\mathbf{z}-\boldsymbol{\mu})^{H}\boldsymbol{\Omega}^{-1}(\mathbf{z}-\boldsymbol{\mu})}.
\end{equation*}

The remaining part of this paper is organized as follows. Section \ref{Problem} formulates the cell-free massive MIMO detection problem. The distributed EP detector is proposed in Section \ref{distributed_EP} and  an analytical framework is provided in Section \ref{SE}. Numerical results are then presented in Section \ref{simulation} and Section \ref{con} concludes the paper.
\section{System Model}\label{Problem}
In this section, we first present the system model and formulate the cell-free massive MIMO detection problem. Then, four commonly-adopted receivers  are briefly introduced.
\subsection{Cell-Free Massive MIMO}
\begin{figure}
  \centering
  \includegraphics[width=3.0in]{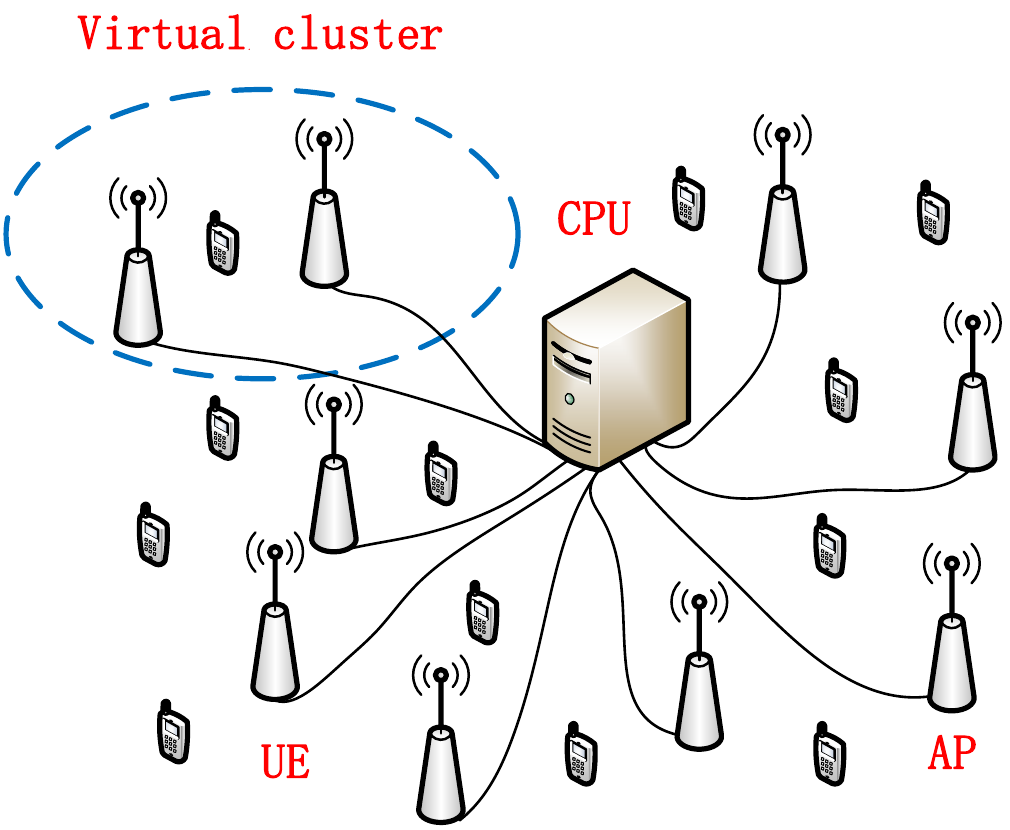}
  \caption{.~~A typical cell-free massive MIMO system.}\label{cell_free}
\end{figure}

As illustrated in Fig.\,\ref{cell_free}, we consider a cell-free massive MIMO network with $L$ distributed APs, each equipped with $N$ antennas to serve $K$ single-antenna UEs. The system can be AP-centric or user-centric.
All  APs are connected to a CPU that has a high computational capability. Denote $\bh_{kl} \sim \mathcal{N}_{\bbC}(\mathbf{0}, \bR_{kl})$ as the channel between the $k$-th user and the $l$-th AP, where $\bR_{kl}\in \bbC^{N\times N}$ is the spatial correlation matrix and $\beta_{k,l} = \mathrm{tr}(\bR_{kl})/N$ is the large-scale fading coefficient involving the geometric path loss and shadowing.  In the uplink data transmission phase, we consider $\mathcal{M}_{k} \subset \{1,\ldots,L\}$ as the subset of APs that serve the $k$-th UE and define the DCC matrices $\bD_{kl}$ based on $\mathcal{M}_{k}$ as
\begin{equation}\label{eqDCC}
  \bD_{kl} = \left\{
  \begin{aligned}
  &\bI_{N} \quad  \mathrm{if} \ l \in \mathcal{M}_{k}   \\
  & \mathbf{0}_{N\times N} \quad  \mathrm{if} \ l \notin \mathcal{M}_{k},
  \end{aligned} \right.
\end{equation}
which is constructed by the DCC strategy.
Furthermore, we define $\mathcal{D}_{l}$ as the set consisting of the UE indices that are served by $l$-th AP :
\begin{equation}\label{eqDl}
  \mathcal{D}_{l} = \bigg\{k: \mathrm{tr}(\bD_{kl}) \geq 1, k\in \{1,\ldots,K \} \bigg\},
\end{equation}
where the cardinality of $\mathcal{D}_{l}$ is denoted as $|\mathcal{D}_{l}|$. When each AP serves all users,
$\mathcal{D}_{l} = K$ accordingly. We assume that perfect channel state information (CSI) is available at the local APs. Therefore, the received signal at the $l$-th AP is given by,
\begin{equation}\label{eqyd}
\by_{l} = \sum_{k=1}^{|\mathcal{D}_{l}|}\sqrt{p_k}\bh_{kl}x_{k}+\bn_{l},
\end{equation}
where $x_{k}\in \bbC$ is the transmitted symbol drawn from the $M$-QAM constellation, and $p_{k}>0$ is the transmit power at the $k$-th UE. The additive noise at the $l$-th AP is denoted as $\bn_{l}\sim \mathcal{N}_{\bbC}(\mathbf{0}_{N},\sigma^{2}\bI_{N})$. Let $\bx=[x_{1},\ldots,x_{K}]$ denote the transmitted vector from all UEs and $\bh_{l}=[\bh_{1l},\ldots,\bh_{Kl}]^{T} \in \bbC^{N \times K}$ as the channel  of the AP $l$ to all UEs. If the $k$-th UE is not associated with the $l$-th AP, the channel vector $\bh_{kl} = \mathbf{0}$ accordingly.
Furthermore, we denote  $\bH=[\bh_{k}^{T},\ldots,\bh_{K}^{T}]^{T}\in\bbC^{LN \times K}$ as the channel matrix between all UEs and APs. The uplink detection problem for cell-free massive MIMO is to detect the transmitted data $\bx$ based on the received signals $\by_{l}\,(l=1,2,\ldots,L)$, channel matrix $\bH$, and noise power $\sigma^{2}$.
\subsection{Linear Receivers}
For the cell-free massive MIMO receiver design, four receivers with different levels of cooperation among APs were introduced and  compared in terms of SE \cite{cell_free_distributed_receiver}.
\begin{itemize}
  \item Level $4$ is a fully centralized receiver where the pilot and data signals received at all APs are sent to the CPU for channel estimation and data detection.
  \item Level $3$ involves two stages. First, each AP estimates the channels and uses the linear MMSE detector to detect the received signals. Then, the detected signals are collected at the CPU for joint detection for all UEs by utilizing the large-scale fading decoding (LSFD) method. Compared to Level $4$, only the channel statistics are utilized at the CPU but the pilot signals are not required to be sent to the CPU.
  \item Level $2$ is a special case of Level $3$. The CPU performs joint detection for all UEs by simply taking the average of the local estimates. Thus, no channel statistics are required to be transmitted to CPU via the fronthaul.
  \item Level $1$ is a fully distributed approach in which the data detection is performed  at the APs based on the local channel estimates. No information is required to be transferred to the CPU.
\end{itemize}

Although the aforementioned four receivers have low complexity, the performance is far from optimal due to the linear structure. This is because all of the four receivers are linear receivers. In contrast, non-linear receivers have shown great advantages in terms of BER while the main limitation is the high computational complexity \cite{EP_detector}. Thanks to the relatively high computing abilities at the CPU and the large number of APs in cell-free massive MIMO systems, we can offload parts of the computational-intensive operations to the CPU and  distribute the partial computation tasks to APs. Next, we will propose a distributed non-linear detector for cell-free massive MIMO systems, which takes advantage of this idea.
\begin{figure}
  \centering
  \includegraphics[width=3.5in]{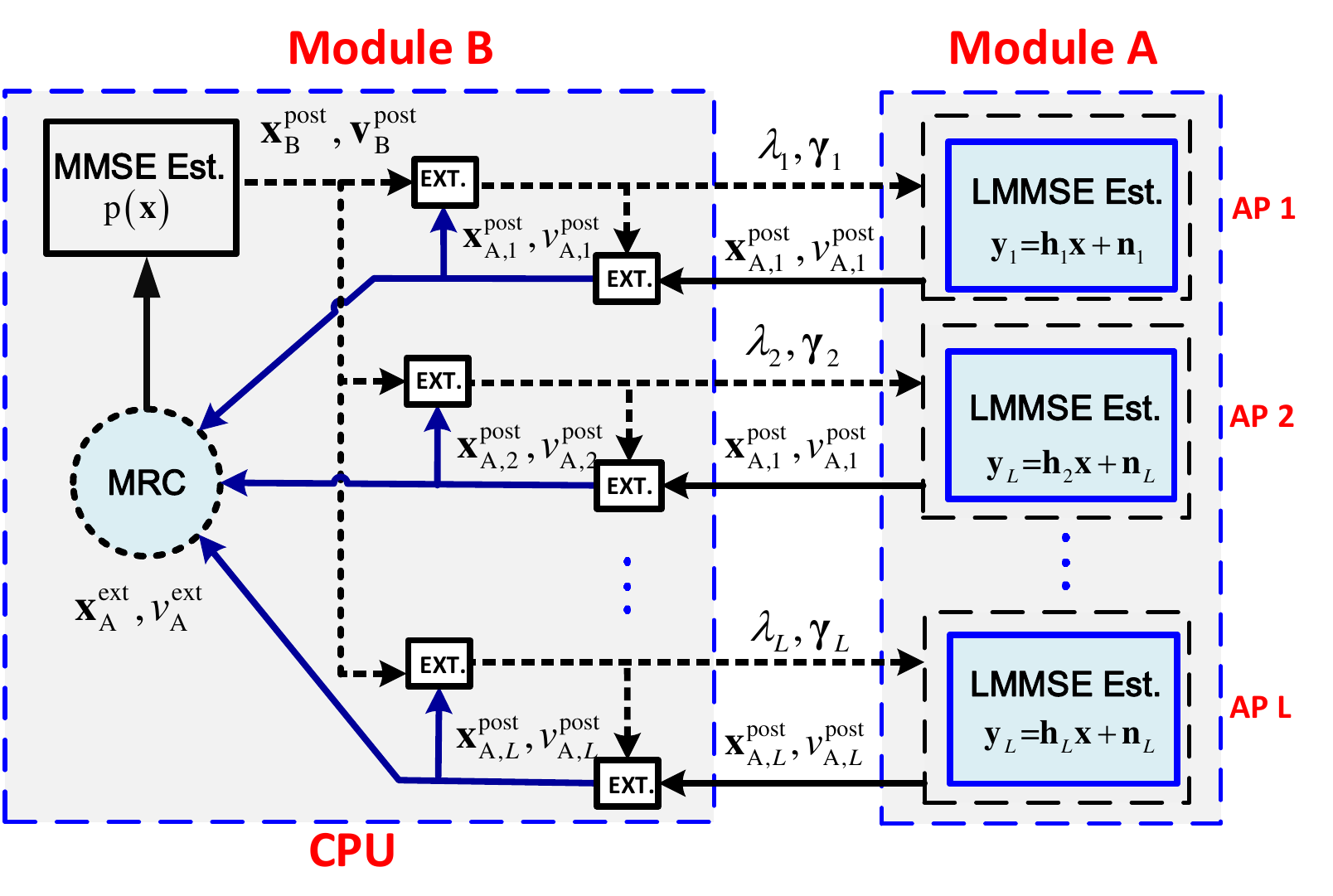}
  \caption{.~~Block diagram of the proposed distributed EP detector. The ``EXT.'' blocks represents the extrinsic information computation.}\label{DeEP}
\end{figure}
\section{Proposed Distributed EP Detector}\label{distributed_EP}
In this section, we apply the EP principle to develop a powerful distributed  MIMO detector for cell-free massive MIMO systems. After introducing the iterative process of the algorithm, we will analyze the computational complexity and fronthaul overhead of the proposed detector.
\subsection{Distributed Bayesian MIMO Detector}
We first utilize the Bayesian inference to recover the signals $\bx$ from the received signal $\by$ in the data detection stage, which is related to the following linear model $\by = \bh \bx+\bn$. Based on Bayes' theorem, the posterior probability is given by
\begin{equation}\label{eq4}
    \mathtt{P}(\bx|\by,\bH)=\frac{\mathtt{P}(\by|\bx,\bH)\mathtt{P}(\bx)}{\mathtt{P}(\by|\bH)}
 = \frac{\mathtt{P}(\by|\bx,\bH)\mathtt{P}(\bx)}{\int \mathtt{P}(\by|\bx,\bH)\mathtt{P}(\bx)d\bx}.
\end{equation}
Given the posterior probability $\mathtt{P}(\bx|\by,\bH)$, the Bayesian MMSE estimate is obtained by
\begin{equation}\label{MMSE_estimate}
\hat{\bx}=\int \bx \mathtt{P}(\bx|\by,\bH)d\bx.
\end{equation}
However, the Bayesian MMSE estimator is not computationally tractable because the marginal posterior probability in
(\ref{MMSE_estimate}) involves a high-dimensional integral. The EP algorithm, proposed in \cite{EP_principle}, provides an iterative method to recover the transmitted $\bx$ from the received signal $\by$. It is derived from the factor graph with the messages updated and passed between different pairs of nodes assumed to follow Gaussian distributions. As the Gaussian distribution can be fully characterized by its mean and variance, only mean and variance are required to be calculated and passed.
\begin{algorithm}\label{algGE}
\caption{Distributed EP for cell-free massive MIMO detection} 
{\bf Input:} 
Received signal $\by_{l}$, channel matrix $\bh_{l}$, noise level $\sigma^{2}$. \\
{\bf Output:} 
Recovered signal $\bx_{\rmB}^{\mathrm{post},T}$.\\
{\bf Initialize:}   
$\boldsymbol{\gamma}_{l}^{(0)} \leftarrow \mathbf{0}$, $\lambda_{l}^{(0)} \leftarrow \frac{1}{E_{x}}$ \\
\For{$t = 1,\cdots, T$ }{
\hspace*{0.02in} {\bf Module A in APs:}
(1) Compute the posterior mean and variance of $\bx_{\rmA,l}$:
\begin{equation}\label{eqvarA}
v_{\rmA,l}^{ \mathrm{post}} \leftarrow \boldsymbol{\Sigma}_{l}^{t} = (\sigma^{-2} \bh^{H}_{l}\bh_{l}+\lambda_{l}^{(t-1)}\bI)
\end{equation}
\begin{equation}\label{eqmeanA}
\bx_{\rmA,l}^{\mathrm{post}} \leftarrow \boldsymbol{\mu}_{l}^{t} = \boldsymbol{\Sigma}_{l}^{t} \bigg( \sigma^{-2}\bh_{l}\by_{l}+
\boldsymbol{\gamma}_{l}^{(t-1)}\bigg)
\end{equation}
\hspace*{0.02in} {\bf Module B in CPU:}
(2) Compute the extrinsic mean and variance of $\bx_{\rmA,l}$:
    \begin{equation}\label{eqextvar}
     v_{\rmA,l}^{\mathrm{ext}} \leftarrow \bigg( \frac{1}{v_{\rmA,l}^{ \mathrm{post}}}-\lambda_{l}^{(t-1)}
      \bigg)^{-1}
    \end{equation}
\begin{equation}\label{eqextmean}
     \bx_{\rmA,l}^{\mathrm{ext}} \leftarrow v_{\rmA,l}^{\mathrm{ext}} \bigg(     \frac{\boldsymbol{\mu}_{l}^{t}}{v_{\rmA,l}^{ \mathrm{post}}}-\boldsymbol{\gamma}_{l}^{(t-1)}
      \bigg)^{-1}
\end{equation}
(3) MRC combining of $\bx_{\rmA,l}$:
\begin{equation}\label{eqvarmrc}
    \frac{1}{v_{\rmA}^{\mathrm{ext}}} = \sum_{l=1}^{L} \frac{1} {v_{\rmA,l}^{\mathrm{ext}}}
\end{equation}
\begin{equation}\label{eqmeanmrc}
    \bx_{\rmA}^{\mathrm{ext}} = v_{\rmA}^{\mathrm{ext}}\sum_{l=1}^{L}\frac{\bx_{\rmA,l}^{\mathrm{ext}}}{v_{\rmA,l}^{\mathrm{ext}}}
\end{equation}
(4) Compute the posterior mean and variance of $\bx_{\rmB}$:
    \begin{equation}\label{eqmeanB}
    \bx_{\rmB}^{\mathrm{post}} = \mathtt{E} \{\bx|\bx_{\rmA}^{\mathrm{ext}}, v_{\rmA}^{\mathrm{ext}}\}
  \end{equation}
   \begin{equation}\label{eqvarB}
    \bv_{\rmB}^{\mathrm{post}} = \mathtt{var} \{\bx|\bx_{\rmA}^{\mathrm{ext}}, v_{\rmA}^{\mathrm{ext}}\}
   \end{equation}
(5) Compute the extrinsic mean and variance $\bx_{\rmB,l}$:
\begin{equation}\label{eqvarest}
    \frac{1}{v_{\rmB,l}^{\mathrm{ext}}} \leftarrow \lambda_{l}^{(t)}
    = \frac{1}{\mathrm{mean}(\bv_{\rmB}^{\mathrm{post}})}-\frac{1}{v_{\rmA,l}^{\mathrm{ext}}}
\end{equation}
\begin{equation}\label{eqmmeanest}
  \frac{\bx_{\rmB,l}^{\mathrm{ext}}}{v_{\rmB,l}^{\mathrm{ext}}} \leftarrow \boldsymbol{\gamma}_{l}^{(t)}=    \frac{\bx_{\rmB}^{\mathrm{post}}}{\mathrm{mean}(\bv_{\rmB}^{\mathrm{post}})} -
  \frac{\bx_{\rmA,l}^{\mathrm{post}}}{v_{\rmA,l}^{\mathrm{ext}}}
\end{equation}
}
\end{algorithm}
Different from the conventional EP-based \cite{EP_detector}, the posterior probability in (\ref{eq4}) has to be rewritten in a distributed way as follows,
\begin{equation}\label{eqdistributed}
  \mathtt{P}(\bx|\by,\bH) \varpropto \mathtt{P}(\bx)\prod_{l=1}^{L} \mathrm{exp}(-\|\by_{l}-\bh_{l}\bx\|^{2}/\sigma^{2}).
\end{equation}
Leveraging the computational capability in the AP, we can deploy partial calculation modules of the EP detector at the AP based on the local information and then send the posterior mean and variance estimates to the CPU for combining. The distributed EP-based detector is illustrated in \textbf{Algorithm 1}. The input of the algorithm is the received signal $\by_{l}$, channel matrix $\bh_{l}$, and noise level $\sigma^{2}$ while the output is the recovered signal $\bx_{\rmB}^{\mathrm{post},T}$ in the $T$-th iteration. The initial parameters are $\boldsymbol{\gamma}_{l}^{(0)} = \mathbf{0}$, $\lambda_{l}^{(0)} = \frac{1}{E_{x}}$, where 
\begin{equation}\label{eqex}
E_{x}=\mathtt{E}\{\|\bx\|^{2}\}/K.
\end{equation}
$\boldsymbol{\gamma}_{l}^{(0)}$ and $\lambda_{l}^{(0)}$ are the initialized extrinsic information and $E_{x}$ is the power of the transmitted symbol $x_{k}$. Furthermore, the block diagram of the proposed distributed detector is illustrated in Fig.\,\ref{DeEP}, which is composed of module A and module B. Each module uses the turbo principle in  iterative decoding. That is, each module passes the extrinsic messages to its next module and  this process is executed until convergence.

To better understand the distributed EP detection algorithm, we elaborate the details for each iteration in \textbf{Algorithm 1}. Specifically, module A is the linear MMSE (LMMSE) estimator performed at the APs according to the following linear model
\begin{equation}\label{eqLMMSE}
  \by_{l} = \bh_{l}\bx+\bn_{l}.
\end{equation}
In the $t$-iteration of the algorithm, the explicit expression for the posterior covariance matrix $\boldsymbol{\Sigma}_{l}^{t}$ and mean vector $\boldsymbol{\mu}_{l}^{t}$  are given by (\ref{eqvarA}) and (\ref{eqmeanA}), respectively. Note that each AP only uses the local channel $\bh_{l}$ to detect the transmitted signal $\bx$. For ease of notation, we omit the iteration index $t$ for all mean and variance estimates. Then, the variance $v_{\rmA,l}^{\mathrm{post}} = \mathrm{tr}(\boldsymbol{\Sigma}_{l}^{t})/|\mathcal{D}_{l}|$ and mean estimate $\bx_{\rmA,l}^{\mathrm{post}} = \boldsymbol{\mu}_{l}^{t}$ are transferred to the CPU to compute the extrinsic information $v_{\rmA,l}^{\mathrm{ext}}$ (\ref{eqextvar}) and  $\bx_{\rmA,l}^{\mathrm{ext}}$ (\ref{eqextmean}), respectively. The extrinsic information $\bx_{\rmA,l}^{\mathrm{\mathrm{ext}}}$ can be regarded as the AWGN observation given by
\begin{equation}\label{eqAWGNl}
  \bx_{\rmA,l}^{\mathrm{ext}} = \bx+\bn^{\mathrm{eq}}_{l},
\end{equation}
where $\bn^{\mathrm{eq}}_{l}\sim \mathcal{N}_{\bbC} (0,v_{\rmA,l}^{\mathrm{ext}}\bI)$ \cite{EP_distributed}. Therefore, the linear model in (\ref{eqyd}) is decoupled into $K$ parallel and independent AWGN channels with equivalent noise $v_{\rmA,l}^{\mathrm{ext}}$. Subsequently, the CPU collects all extrinsic means $\{\bx_{\rmA,l}^{\mathrm{ext}}\}_{l=1}^{L}$ and variances  $\{v_{\rmA,l}^{\mathrm{ext}}\}_{l=1}^{L}$ and performs MRC. The MRC expressions (\ref{eqvarmrc}) and (\ref{eqmeanmrc}) are obtained by maximizing the post-combination signal-to-noise ratio (SNR) of the final AWGN observation $\bx_{\rmA}^{\mathrm{ext}}$  at the CPU, given by
\begin{equation}\label{eqAWGN}
  \bx_{\rmA}^{\mathrm{ext}} = \bx+\bn^{\mathrm{eq}},
\end{equation}
where $\bn_{\rmA}^{\mathrm{ext}} \sim \mathcal{N}_{\bbC} (0,v_{\rmA}^{\mathrm{ext}}\bI)$. The CPU uses the posterior mean estimator to detect the signal $\bx$ from the equivalent AWGN model (\ref{eqAWGN}). Then,
the posterior mean and variance are computed by the posterior MMSE estimator for the equivalent AWGN model in (\ref{eqmeanB}) and (\ref{eqvarB}).
As the transmitted symbol is assumed drawn from the $M$-QAM set $\mathcal{S}=\{s_{1}, s_{2}, \ldots, s_{M}\}$, the corresponding expressions for each element in (\ref{eqmeanB}) and (\ref{eqvarB}) are given by
\begin{equation}\label{eqmean}
  [x_{\rmB}^{\mathrm{post}}]_{k} = \frac{\sum_{s_{i} \in \mathcal{S}}s_{i}\mathcal{N}_{\bbC}(s_{i};[x_{\rmA}^{\mathrm{ext}}]_{k}, v_{\rmA}^{\mathrm{ext}})p(s_{i})}{\sum_{s_{i} \in \mathcal{S}}\mathcal{N}_{\bbC}(s_{i};[x_{\rmA}^{\mathrm{ext}}]_{k}, v_{\rmA}^{\mathrm{ext}})p(s_{i})}
\end{equation}

\begin{equation}\label{eqmean}
  [v_{\rmB}^{\mathrm{post}}]_{k} = \frac{\sum_{s_{i} \in \mathcal{S}}|s_{i}|^{2}\mathcal{N}_{\bbC}(s_{i};[x_{\rmA}^{\mathrm{ext}}]_{k}, v_{\rmA}^{\mathrm{ext}})p(s_{i})}{\sum_{s_{i} \in \mathcal{S}}\mathcal{N}_{\bbC}(s_{i};[x_{\rmA}^{\mathrm{ext}}]_{k}, v_{\rmA}^{\mathrm{ext}})p(s_{i})}
  -|[x_{\rmB}^{\mathrm{post}}]_{k}|^{2}.
\end{equation}
where $[x_{\rmB}^{\mathrm{post}}]_{k}$, $[v_{\rmB}^{\mathrm{post}}]_{k}$, and $[x_{\rmA}^{\mathrm{ext}}]_{k}$ are the $k$-th element in $\bx_{\rmB}^{\mathrm{post}}$, $\bv_{\rmB}^{\mathrm{post}}$, and $[\bx_{\rmA}^{\mathrm{ext}}]_{k}$, respectively. The posterior mean and variance $\bx_{\rmB}^{\mathrm{post}}$ and $\bv_{\rmB}^{\mathrm{post}}$ are then utilized to compute the extrinsic information $\lambda_{l}^{(t)}$ and $\boldsymbol{\gamma}_{l}^{(t)}$ for each AP in (\ref{eqvarest}) and (\ref{eqmmeanest}), where the function $\mathrm{mean}(\cdot)$ is used to compute the mean. Finally, the extrinsic information $\lambda_{l}^{(t)}$ and $\boldsymbol{\gamma}_{l}^{(t)}$ are transferred to each AP in the next iteration. The whole procedure is executed iteratively until terminated by a certain stopping criterion or a maximum number of iterations.
\begin{table}[h]
	\centering
	\renewcommand{\arraystretch}{1.1}
	\begin{minipage}[c]{1\columnwidth}
		\centering
		\caption{.~~Complexity of different detectors}
		\label{tbl:complexity}
		\begin{tabular}{@{}lcccc@{}}
			\toprule
			Detectors & Distributed EP & Level 4 & Level 3\&2 & Level 1   \\
			\midrule
			  AP & $O(T|\mathcal{D}_{l}|N^{2})$ & $0$& $O(|\mathcal{D}_{l}|N^{2})$&$O(|\mathcal{D}_{l}|N^{2})$ \\	
			\midrule
              CPU & $O(T|\mathcal{D}_{l}|^{2})$ & $O(LN)^{3}$& $O(|\mathcal{D}_{l}|)$&$0$ \\						
			\bottomrule
		\end{tabular}
	\end{minipage}
\end{table}

\subsection{Computational Complexity and Fronthaul Overhead}
In the following, we provide the complexity analysis for different detectors in Tables \ref{tbl:complexity} and \ref{tbl:fronthaul}.
For the proposed distributed EP detector, the computational complexity at each AP is dominated by the LMMSE estimator for estimating the signal $\bx_{\mathrm{A},l}^{\mathrm{post}}$, which is $O(|\mathcal{D}_{l}|^{3})$ because of the matrix inversion is required  in (\ref{eqvarA}) while the computational complexity at the CPU is $O(|\mathcal{D}_{l}|^{2})$ in each iteration. Furthermore, if the number of antennas $N$ is less than $|\mathcal{D}_{l}|$, we can use the matrix inversion lemma to carry out the matrix inversion in (\ref{eqvarA}) as follows
\begin{align}\label{eqmatrixinv}
  & (\sigma^{-2}\bh_{l}^{H}\bh_{l}+\bD)^{-1} = \bD^{-1} \\ \nonumber
  &-\sigma^{-2}\bD^{-1}\bh_{l}^{H}(\bI+\sigma^{-2}\bh_{l}\bD^{-1}\bh_{l}^{H})^{-1}\bh_{l}\bD^{-1},
\end{align}
where $\bD = \lambda_{l}^{(t-1)}\bI$ and  the computational complexity is reduced to $O(|\mathcal{D}_{l}| N^{2})$. Therefore, the overall computational complexity at each AP is $O(T|\mathcal{D}_{l}|N^{2})$ while
the overall computational complexity at the CPU is $O(T|\mathcal{D}_{l}|^{2})$ for $T$ iterations. As can be observed in Table \ref{tbl:complexity}, the distritbuted EP detector mainly increases the computational complexity at the CPU when compared with other distributed linear receivers (Levels 1-3).
\begin{table}[h]
	\centering
	\renewcommand{\arraystretch}{1.1}
	\begin{minipage}[c]{1\columnwidth}
		\centering
		\caption{.~~Fronthaul overhead}
		\label{tbl:fronthaul}
		\begin{tabular}{@{}lcc@{}}
		  \toprule
		   Detectors & Coherence block &  Statistical parameters  \\
		  \midrule
		  Distributed EP  &$\sum\limits_{l=1}^{L}(\tau_{c}-\tau_{p})2T(|\mathcal{D}_{l}|+1)$&$0$  \\
		  \midrule
		  Level 4  &$\tau_{c}NL$ &$KLN^{2}/2$ \\
		  \midrule
		  Level 3  &$(\tau_{c}-\tau_{p})KL$&$KL+(L^{2}K^{2}+KL)/2$  \\
		  \midrule
		  Level 2  &$(\tau_{c}-\tau_{p})KL$&$0$  \\
		  \midrule
		  Level 1  &$0$&$0$  \\				
		 \bottomrule
		\end{tabular}
	\end{minipage}
\end{table}

We compare the number of complex scalars that need to be transmitted from the APs to the CPU via the fronthauls in Table \ref{tbl:fronthaul}. We assume that $\tau_c$ and $\tau_p$ are the coherence time and pilot length, respectively. The results for the 4 level detectors are from \cite{cell_free_distributed_receiver}. With the proposed detector, $\sum\limits_{l=1}^{L}(\tau_{c}-\tau_{p})2T(|\mathcal{D}_{l}|+1)$ scalars need to be passed from the APs to the CPU and no statistical parameters are required to be passed, where $T$ denotes the total iterations. The fronthaul overhead of the proposed distributed EP detector is similar to Level 4 and the detailed comparison is determined by the value of the system parameters.
\section{State Evolution Analysis}\label{SE}
In this section, we provide an analytical framework to predict the asymptotic performance of the distributed EP in the large system limit. We consider $L,K,N \rightarrow \infty$ and fix
\begin{equation}\label{eqratio}
  \alpha_{l} = \frac{K}{N}, \alpha = \frac{K}{LN}.
\end{equation}
Then, we have following proposition.
\begin{proposition}\label{proposition1}
In the large-system limit, the asymptotic behavior (such as MSE and BER) of  Algorithm 1 can be described by the following equations:
\begin{subequations} \label{state_evolution}
   \begin{align}
   v_{\mathrm{A},l}^{\mathrm{ext},t} &= \frac{\alpha_{l}\sigma^{2}+(\alpha_{l}-1)\lambda_{l}^{(t-1)}}{2} \\ \nonumber
   &+\frac{\sqrt{\bigg(\alpha_{l}\sigma^{2}+(\alpha_{l}-1)\lambda_{l}^{(t-1)}\bigg)^{2}+4\alpha_{l}\sigma^{2}\lambda_{l}^{(t-1)}}}{2}  \\
   &v_{\mathrm{A}}^{\mathrm{ext},t} = \bigg( \sum_{l=1}^{L} \frac{1}{ v_{\mathrm{A},l}^{\mathrm{ext}, t}}\bigg)^{-1} \\
   &\lambda_{l}^{(t)} = \frac{1}{\mathrm{MSE}(v_{\mathrm{A}}^{\mathrm{ext},t})}-\frac{1}{v_{\mathrm{A},l}^{\mathrm{ext},t}}.
   \end{align}
\end{subequations}
\hfill\ensuremath{\blacksquare}
\end{proposition}
The function $\mathrm{MSE}(\cdot)$ is given by
\begin{equation}\label{eqMSE}
\mathrm{MSE}(v_{\mathrm{A},l}^{\mathrm{ext},t}) \triangleq \mathtt{E}\bigg\{| x-\mathtt{E}\{x|x_{\mathrm{A},l}^{\mathrm{ext},t},v_{\mathrm{A},l}^{\mathrm{ext},t}|^{2}\bigg\}
\end{equation}
and the expectation is with respect to $x$. The state equations can be proved using the  method in \cite{Takeuchi}, and we give an intuitive explanation here, which is linked to \textbf{Algorithm 1}. By substituting (\ref{eqvarA}) into (\ref{eqextvar}), we have following expression as
\begin{equation}\label{eqextvarasy}
v_{\rmA,l}^{\mathrm{ext},t} = \frac{1}{\frac{1}{K}\mathrm{tr}(\sigma^{-2} \bh^{H}_{l}\bh_{l}+\lambda_{l}^{(t-1)}\bI)}-\lambda_{l}^{(t-1)}.
\end{equation}

Notably, $v_{\rmA,l}^{\mathrm{ext}}$ can be characterized by the eigenvalues of $\bh^{H}_{l}\bh_{l}$, which converge to a deterministic distribution in the large system limit. Because of the limited space, we only show the final result of the \textbf{Proposition 1} and omit the derived process. By adopting  the result of the $\mathcal{R}$-transform  of the average empirical eigenvalue distribution, we have an asymptotic expression for $v_{\rmA,l}^{\mathrm{ext},t}$ in (\ref{state_evolution}).  Then, the asymptotic expressions for $v_{\mathrm{A}}^{\mathrm{ext},t}$ and $\lambda_{l}^{(t)}$ can be derived from (\ref{eqvarmrc}) and (\ref{eqvarest}).
Finally, the asymptotic MSE can be interpreted as the MSE of the decoupled scalar AWGN channels (\ref{eqAWGN}) and is related to distribution of transmitted signal $\bx$. Next, we will give a specific example for \textbf{Proposition 1}.

{\noindent {\bf Example~1:}} If the data symbol is drawn from a quadrature phase-shift keying (QPSK) constellation, the $\mathrm{MSE}$ is expressed by
\vspace{-0.15in}
\begin{equation}\label{mse_x_explicit expression}
  \mathrm{MSE} = 1-\int \rmD z\tanh(\frac{1}{v_{\mathrm{A},l}^{\mathrm{ext},t}}+\sqrt{\frac{1}{v_{\mathrm{A},l}^{\mathrm{ext},t}}}z).
\end{equation}
Furthermore, the final BER w.r.t. $\bx$ can also be evaluated through the equivalent AWGN channel (\ref{eqAWGN}) with an equivalent $\mathrm{SNR} = 1/v_{\mathrm{A},l}^{\mathrm{ext},T}$ and is given by \cite{HeJSTSP}
\begin{equation}\label{SER_QPSK}
  \mathrm{BER}= 2Q(\sqrt{\frac{1}{v_{\mathrm{A},l}^{\mathrm{ext},T}}})-[Q(\sqrt{\frac{1}{v_{\mathrm{A},l}^{\mathrm{ext},T}}})]^{2},
\end{equation}
where $Q(x)= \int_{x}^{\infty}\rmD z$ is the $Q$-function. In fact, the MSE and BER are determined on the basis of the knowledge of the  AWGN channel (\ref{eqAWGN}) with $\mathrm{SNR} = 1/v_{\mathrm{A},l}^{\mathrm{ext},T}$, which is known as the decoupling principle. Thus, if the data symbol is drawn from other $M$-QAM constellations, then the corresponding BER can be easily obtained using the closed-form BER expression in \cite{Proakis2007}. Note that the state evolution equations are accurate when each AP serves all UEs in the network and the channel matrix $\bH$ is a Rayleigh fading channel. This assumption can be achieved by perfect power control to compensate the large-scaled fading effect.
\section{Simulation Results}\label{simulation}
In this section, we provide sample simulation results to demonstrate the performance of the proposed distributed EP detector for cell-free massive MIMO.  Consider $L=16$, $N=8$, $K=16$, and the other simulation parameters are the same as those in \cite{cell_free_distributed_receiver}. The SNR is defined as $\mathrm{SNR} = 1/\sigma^{2}$.
\begin{figure}[h]
  \centering
  \includegraphics[width=3in]{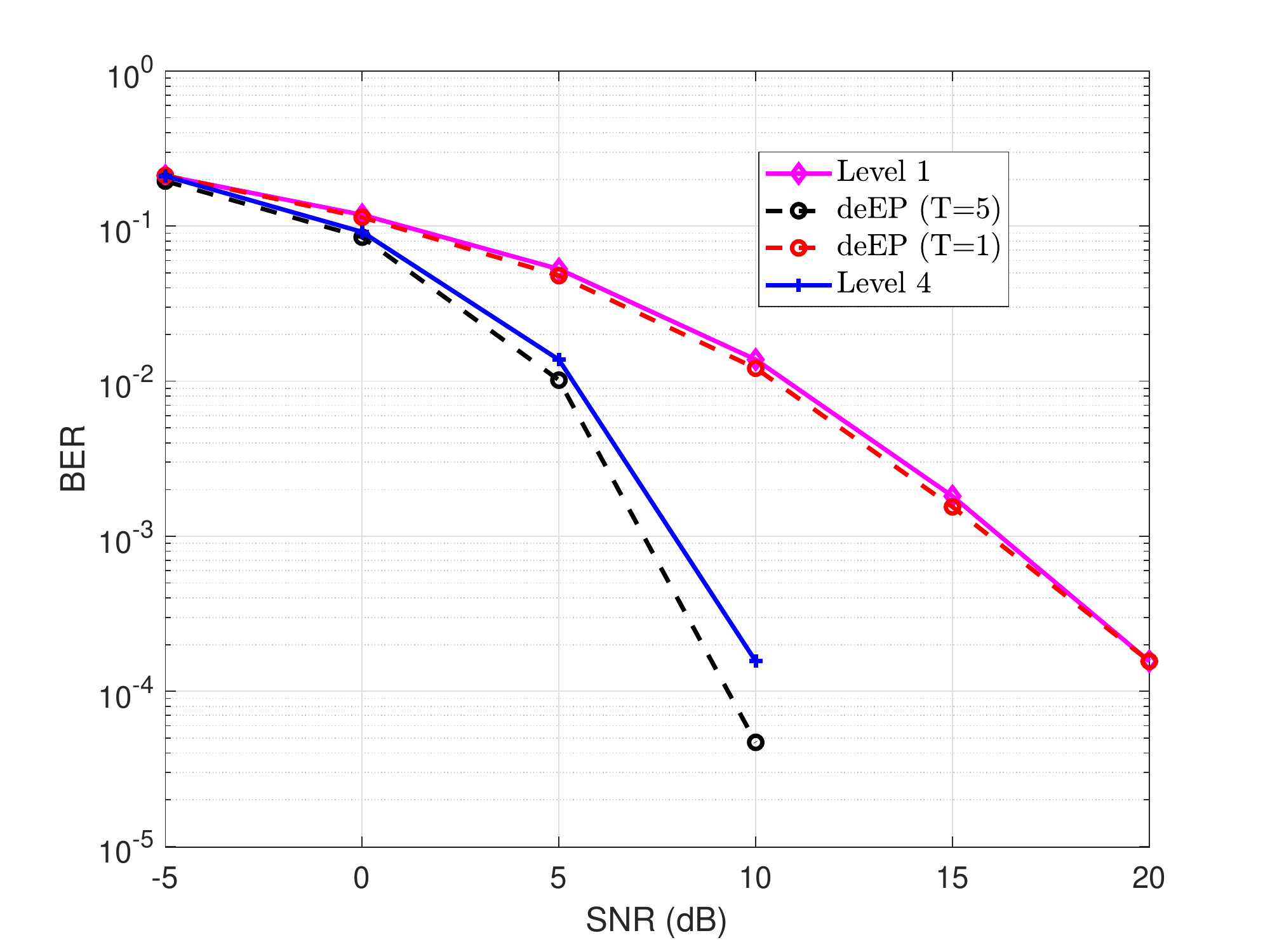}
  \caption{.~~BER performance comparisons of different detectors in the original cell-free massive MIMO system.}\label{Fig:nonscalable}
\end{figure}
\subsection{Original Cell-Free Massive MIMO}
In this subsection, we consider the BER of the proposed detector in the original cell-free massive MIMO where each AP serves all UEs. We assume that perfect CSI can be obtained at each AP. Fig.\,\ref{Fig:nonscalable} compares the achievable BER of the proposed distributed EP detector with other detectors investigated in \cite{cell_free_distributed_receiver}. The results are obtained by the Monte Carlo simulation with $10,000$ independent channel realizations. We denote ``deEP" as the distributed EP detector. It can be observed that the proposed distributed EP detector is comparable with the Level 1 detector with only one EP iteration and outperforms the centralized Level 4 detector with $T=5$ iterations.

\begin{figure}
  \centering
  \includegraphics[width=3in]{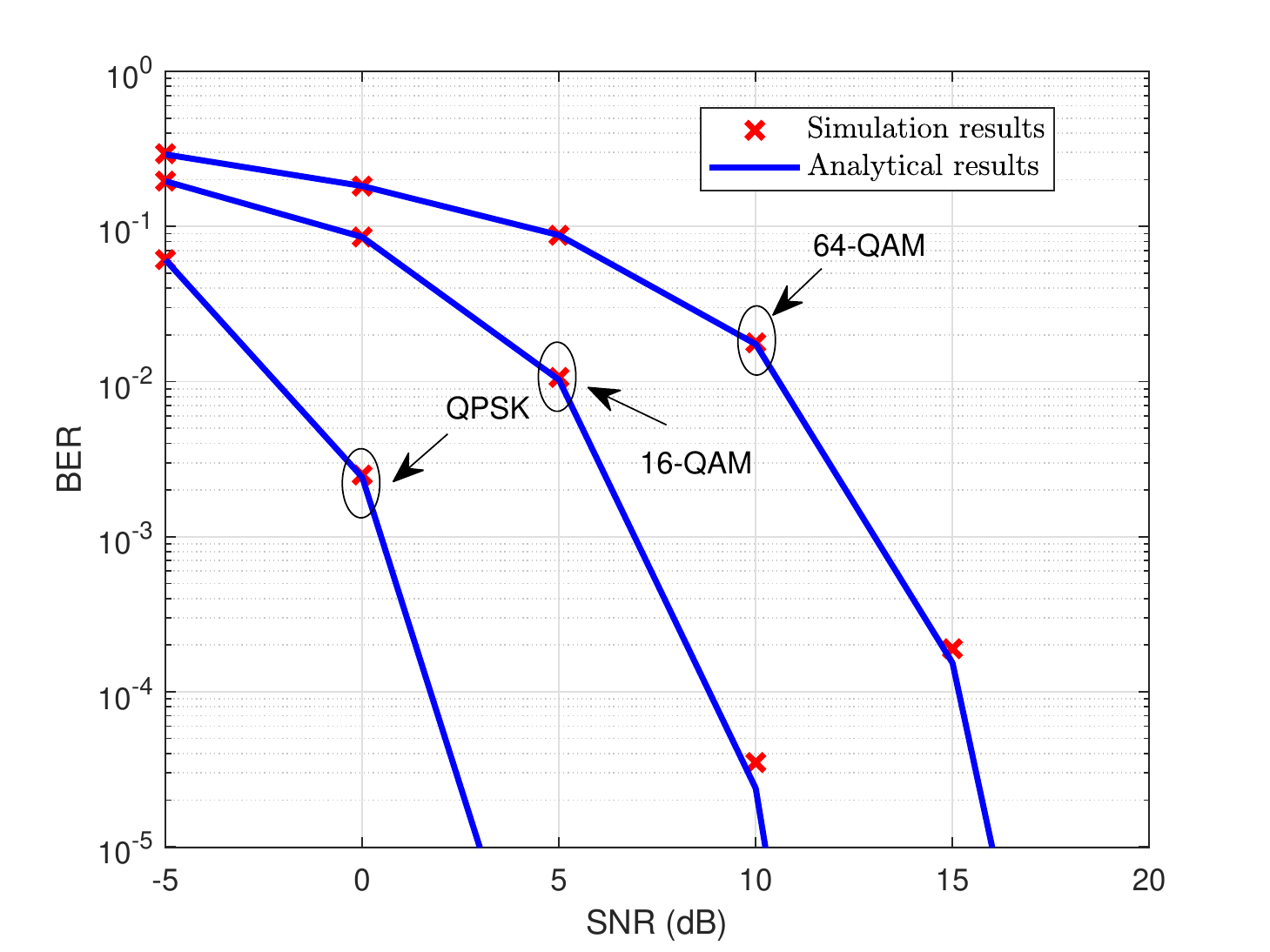}
  \caption{.~~BER performance comparisons of the analytical and simulation results of the distributed EP detector with different modulation schemes.}\label{Fig:analytical}
\end{figure}

We next the accuracy of the analytical framework in Fig.\,\ref{Fig:analytical} with different modulation schemes. As shown in the figure, the BERs of the proposed detector match well with the derived analytical results, which demonstrate the accuracy of the analytical framework. Therefore, instead of performing time-consuming Monte Carlo simulations to obtain the corresponding performance metrics, we can predict the theoretical behavior by state  equations. Furthermore, the analytical framework can be further utilized to optimize the system design.
\subsection{Scalable Cell-Free Massive MIMO}
\begin{figure}
  \centering
  \includegraphics[width=3in]{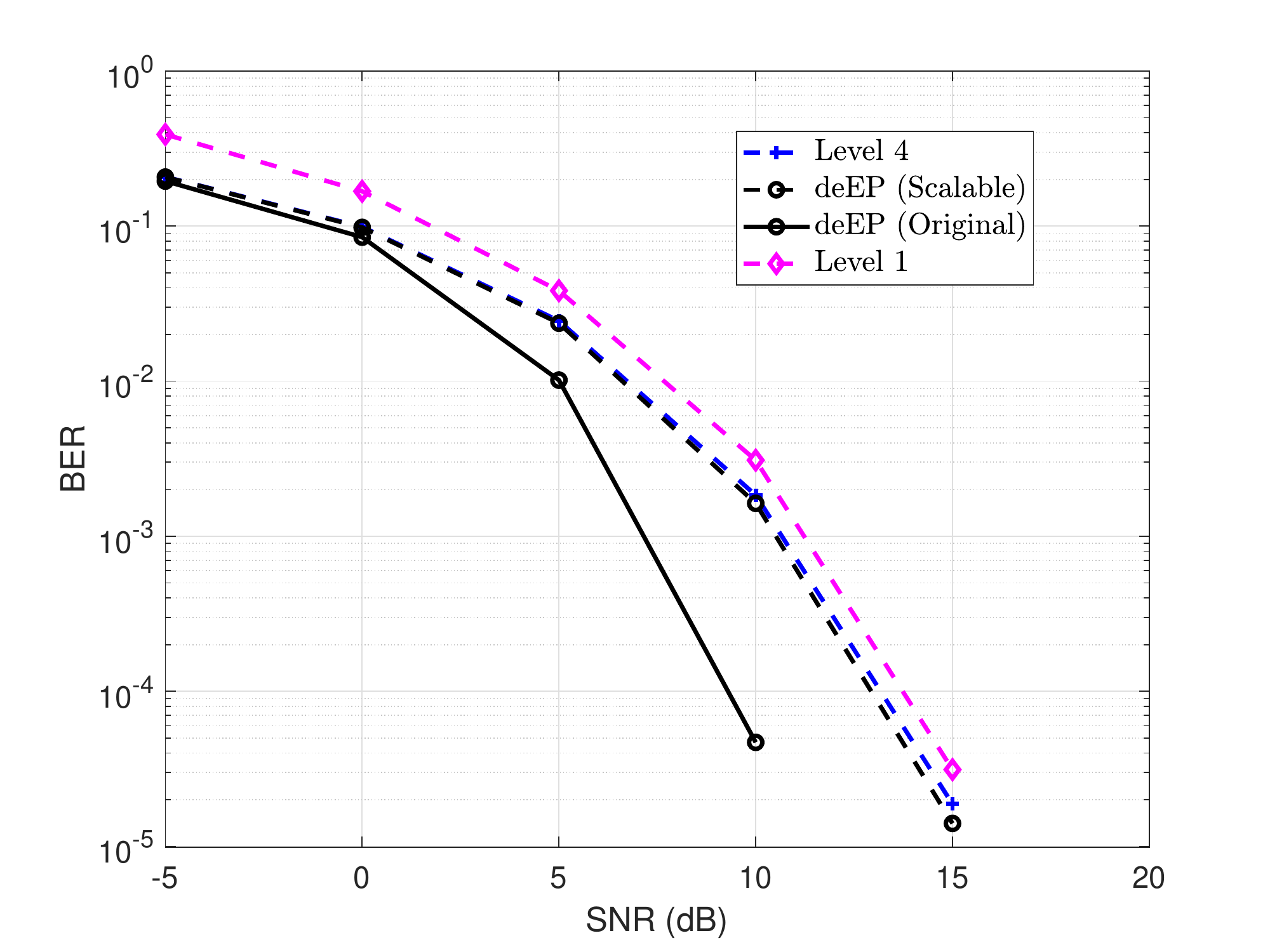}
  \caption{.~~BER performance comparisons of different detectors in cell-free massive MIMO with dynamic cluster cooperation.}\label{Fig:DCC}
\end{figure}

As the \emph{user-centric} approach is more attractive for cell-free massive MIMO, we then investigate the distributed EP detector with DCC. The accessing UE first appoints a master AP according to the large-scale fading factor and assigns a pilot to the appointed AP. Then, other neighboring APs determine whether they serve the accessing UE according to the assigned pilot. Finally, the cluster for the $k$-th UE  is constructed. Fig.\,\ref{Fig:DCC} shows that the distributed EP detector outperforms both the centralized and distributed MMSE detectors. Furthermore, the performance loss is acceptable when compared to original cell-free massive MIMO and the computational complexity is significantly decreased from $O(KN^{2})$ to $O(|\mathcal{D}_{l}|N^{2})$.
\section{Conclusion}\label{con}
In this paper, we proposed a distributed EP detector for cell-free massive MIMO. It is shown that such detector can achieve better performance than other linear receivers for both original and scalable cell-free massive MIMO networks. Compared to other distributed detectors, it can achieve a better BER performance and mainly increase the computational overhead at the CPU. An analytical framework was also provided to describe the asymptotic performance of the proposed detector in a large system setting. Simulation results have also been provided and demonstrated  that the proposed method outperforms the existing distributed detectors for cell-free massive MIMO in terms of BER performance.


\begin{thebibliography}{10}


\bibitem{massiveMIMO}
T. L. Marzetta, ``Noncooperative cellular wireless with unlimited numbers
of base station antennas,'' {\em IEEE Trans. Wireless Commun.}, vol. 9,
no. 11, pp. 3590--3600, Nov. 2010.

\bibitem{cell_free}
H. Q. Ngo, A. Ashikhmin, H. Yang, E. G. Larsson, and T. L. Marzetta,
''Cell-free massive MIMO versus small cells,'' {\em IEEE Trans. Wireless
Commun.}, vol. 16, no. 3, pp. 1834--1850, Mar. 2017.

\bibitem{cell_free_Zhang}
J. Zhang, S. Chen, Y. Lin, J. Zheng, B. Ai, and L. Hanzo, ``Cell-free massive MIMO: A new next-generation paradigm,'' {\em IEEE Access}, vol. 7, pp. 99878--99888, Sep. 2019.

\bibitem{6G}
K. B. Letaief, W. Chen, Y. Shi, J. Zhang, and Y.-J.-A. Zhang, ``The
roadmap to 6G: AI empowered wireless networks,'' {\em IEEE Commun.
Mag.}, vol. 57, no. 8, pp. 84-90, Aug. 2019.


\bibitem{DCC}
E. Bj\"{o}rnson, N. Jald\'{e}n, M. Bengtsson, and B. Ottersten, ``Optimality
properties, distributed strategies, and measurement-based evaluation
of coordinated multicell OFDMA transmission,'' {\em IEEE Trans. Signal
Process.}, vol. 59, no. 12, pp. 6086--6101, Dec. 2011.

\bibitem{scalable}
E. Bj\"{o}rnson and L. Sanguinetti, ``Scalable cell-free massive MIMO
systems,'' {\em IEEE Trans. Commun.}, vol. 68, no. 7, pp.
4247--4261, Jul. 2020.



\bibitem{LSFD}
E. Nayebi, A. Ashikhmin, T. L. Marzetta, and B. D. Rao, ``Performance of cell-free massive MIMO systems with MMSE and LSFD receivers,'' in {\em Proc. 50th Asilomar Conf. Signals, Syst. Comput.}, Nov. 2016, pp. 203--207.

\bibitem{cell_free_distributed_receiver}
E. Bj\"{o}rnson and L. Sanguinetti, ``Making cell-free massive MIMO competitive with MMSE processing and centralized
implementation,'' {\em IEEE Trans. Wireless Commun.}, vol. 19, no. 1, pp. 77--90, Jan. 2020.

\bibitem{EP_principle}
T. P. Minka, ``A family of algorithms for approximate Bayesian Inference,'' Ph.D. dissertation, Dept. Elect. Eng. Comput. Sci., MIT, Cambridge, MA, USA, 2001.

\bibitem{Decentralized-AMP}
C. Jeon, K. Li, J. R. Cavallaro, and C. Studer, ``Decentralized equalization with feedforward architectures for massive MU-MIMO,'' {\em IEEE Trans. Signal Process.}, vol. 67, no. 17, pp. 4418--4432, Sep. 2019.

\bibitem{EP_distributed}
H. Wang, A. Kosasih, C. Wen, S. Jin, and W. Hardjawana, ``Expectation
propagation detector for extra-large scale massive MIMO,'' {\em IEEE Trans.
Wireless Commun.}, vol. 19, no. 3, pp. 2036--2051, Mar. 2020.

\bibitem{Takeuchi}
K. Takeuchi, ``Rigorous dynamics of expectation-propagation-based signal recovery from unitarily invariant measurements,'' {\em IEEE Trans. Inf. Theory.}, vol. 66, no. 1, 368--386, Oct. 2019.

\bibitem{EP_detector}
J. C\'{e}spedes, P. M. Olmos, M. Sánchez-Fernández, and F. Perez-Cruz,
``Expectation propagation detection for high-order high-dimensional
MIMO systems,'' {\em IEEE Trans. Commun.}, vol. 62, no. 8, pp. 2840--2849,
Aug. 2014.


\bibitem{HeJSTSP}
H. He, C.-K. Wen, and S. Jin, ``Bayesian optimal data detector for hybrid
mmWave MIMO-OFDM systems with low-resolution ADCs,'' {\em IEEE J. Sel. Topics Signal Process.}, vol. 12, no. 3, pp. 469--483, Jun. 2018.

\bibitem{Proakis2007}
J.~G. Proakis, \emph{Digital Communications}.\hskip 1em plus 0.5em minus
0.4em\relax Boston, USA: McGraw-Hill Companies, 2007.

\end{thebibliography}
\end{document}